\documentclass[twocolumn]{aastex631}
\shorttitle{Scaling Laws for Solar-stellar Atmospheric Heating}
\shortauthors{Toriumi \& Airapetian}
\graphicspath{{./}{figures/}}

\begin{document}

\title{Universal Scaling Laws for Solar and Stellar Atmospheric Heating}

\correspondingauthor{Shin Toriumi}
\email{toriumi.shin@jaxa.jp}

\author[0000-0002-1276-2403]{Shin Toriumi}
\affiliation{Institute of Space and Astronautical Science, Japan Aerospace Exploration Agency, 3-1-1 Yoshinodai, Chuo-ku, Sagamihara, Kanagawa 252-5210, Japan}

\author[0000-0003-4452-0588]{Vladimir S. Airapetian}
\affiliation{Sellers Exoplanetary Environments Collaboration, NASA Goddard Space Flight Center, Greenbelt, MD, USA}
\affiliation{Department of Physics, American University, Washington, DC, USA}








\begin{abstract}
The Sun and sun-like stars commonly host the multi-million-Kelvin coronae and the 10,000-Kelvin chromospheres. These extremely hot gases generate X-ray and Extreme Ultraviolet emissions that may impact the erosion and chemistry of (exo)planetary atmospheres, influencing the climate and conditions of habitability. However, the mechanism of coronal and chromospheric heating is still poorly understood. While the magnetic field most probably plays a key role in driving and transporting energy from the stellar surface upwards, it is not clear if the atmospheric heating mechanisms of the Sun and active sun-like stars can be described in a unified manner. To this end, we report on a systematic survey of the responses of solar and stellar atmospheres to surface magnetic flux over a wide range of temperatures. By analyzing 10 years of multi-wavelength synoptic observations of the Sun, we reveal that the irradiance and magnetic flux show power-law relations with an exponent decreasing from above- to sub-unity as the temperature decreases from the corona to the chromosphere. Moreover, this trend indicating the efficiency of atmospheric heating can be extended to sun-like stars. We also discover that the power-law exponent has a solar cycle dependence, where it becomes smallest at activity maximum, probably due to the saturation of atmospheric heating. Our study provides observational evidence that the mechanism of atmospheric heating is universal among the Sun and sun-like stars, regardless of age or activity.
\end{abstract}



\section{Introduction}\label{sec:introduction}

Observations suggest that the cool stars of spectral classes later than F5, including the Sun with outer convection envelopes, possess hot outer atmospheres, the corona ($\log{T}>6$), the transition region ($\log{T}=4$--6), and the chromosphere ($\log{T}\sim4$) \citep{2004A&ARv..12...71G}. The coronal temperature is up to two orders of magnitude greater than that of the surface layer, the photosphere. However, two fundamental issues regarding the formation of hot atmospheres remain unsolved. First, it is a mystery how such high temperatures are achieved and maintained despite the radiative and conductive cooling \citep{1977ARA&A..15..363W,2006SoPh..234...41K}. Although it is widely argued that the surface magnetic field is crucial for supplying energy and momentum to the upper layers, the mechanisms of response of the atmospheres to the surface magnetic field have not been clarified yet. Second, young solar-type stars are known to exhibit far hotter coronae than their older siblings, including our Sun \citep{2004A&ARv..12...71G}. However, it is not confirmed whether the heating mechanisms are common.

Ionizing radiations in the form of X-ray and Extreme UV (XUV) from planet-hosting stars have a significant impact on close-in exoplanets. In the Sun, strong concentrations of magnetic field, the active regions, are the primary source of XUV emission and arena of catastrophic outbursts called flares and coronal mass ejections, which may directly disturb the terrestrial environments \citep{2010ARA&A..48..241B,2011SSRv..159...19F,2011LRSP....8....6S,2019LRSP...16....3T}. The planetary atmosphere helps protect surface life from harmful radiation, but strong XUV radiation from active regions and flares may drive ionization and the subsequent escape of the atmospheres, threatening the habitability \citep{2019LNP...955.....L,2020IJAsB..19..136A}. On the other hand, it has been indicated that XUV may have triggered the birth of life by forming biologically relevant molecules in lower planetary atmospheres \citep{2016NatGe...9..452A,2017NatSR...714141A}. Therefore, understanding how stellar atmospheric radiations respond to the surface magnetic field is critical for not only revealing the atmospheric heating but also searching for habitable exoplanets in our neighborhood.

Previous studies have shown that in a wide range of scales from solar active regions to the entire Sun and even the late-type (F, G, K, and M) stars, the X-ray radiation power $F$ has a uniform scaling relationship with a power-law index $\alpha$ slightly above unity with respect to the amount of magnetic flux on the stellar surface $\Phi$, $F\propto\Phi^{\alpha}$ ($\alpha=1.1$--1.8) \citep{1998ApJ...508..885F,2003ApJ...598.1387P,2014MNRAS.441.2361V}. The fact that the same power-law relationship continues over the scales suggests a common physical mechanism of the solar/stellar coronal heating. Specifically, the surface magnetic field is the ultimate source of the heating in both regimes, driven by photospheric convection at various scales. If the heating is caused by energy transport from the surface via magnetohydrodynamic waves and dissipation by small-scale turbulence, its intermittent dynamics develops eddies that cascade down to smaller scales, eventually dissipate, and form power-law (i.e. scale-free) spectra with common indices of 1.5 to 2.4 \citep{1941DoSSR..30..301K,1999PhRvL..83.4662B,2016SSRv..198...47A}. Therefore, the scaling law of X-ray flux, $F\propto\Phi^{\alpha}$, represents one of the most fundamental relationships in astrophysics and has provided the basis for investigating the coronal heating of the Sun and stars.

In this paper, we report on the analysis of the scaling laws between irradiances over the range of the wavelength (i.e., of atmospheres of different temperatures) and photospheric magnetic flux, which sets the ground for understanding the universal atmospheric heating mechanism. The previous attempts to derive the scaling laws were based on limited temperature domains \citep[e.g.,][]{1975ApJ...200..747S,1989ApJ...337..964S,1998ApJ...508..885F,2003ApJ...598.1387P,2014MNRAS.441.2361V}, unlike the present novel study wherein a wide range of temperatures is considered, or, the temperature dependence of scaling laws was examined only with regards to indirect magnetic proxies such as the \ion{Ca}{2} K intensity \citep{1991A&A...252..203R}. More recent studies, such as \citet{2009A&A...497..273L} and \citet{2018A&A...619A...5B}, performed multi-wavelength analysis using solar chromospheric and transition-region lines and continuum and found that the power-law index $\alpha$ increases with the formation temperature. Our study expands these studies and builds a bridge to the stellar regime.

The results presented here are based on the analysis of ``Sun-as-a-star'' synoptic data over 10 years (almost one solar activity cycle), which represents the global behavior of the Sun as a whole (spatially integrated over the disk), and a comparison of the obtained solar power-law relations with currently available stellar data.

\section{Analysis}\label{sec:analysis}

\subsection{Total Unsigned Magnetic Flux of the Sun}

First, we characterized the variation in the unsigned magnetic flux on the visible hemisphere of the Sun. Whereas some of the previous works \citep[e.g.,][]{1989ApJ...337..964S,2009A&A...497..273L} measured the magnetic field strength, which was limited in the magnetic field range (e.g., up to 250 G), we adopted the magnetic flux because it has an advantage of expanding the dynamic range. Also, by using the total unsigned flux for the Sun, we can compare the ``unresolved'' Sun with other sun-like stars by integrating the solar inhomogeneous magnetic flux over the entire hemisphere. For this purpose, we used the sequence of full-disk synoptic magnetograms obtained by the Helioseismic and Magnetic Imager (HMI; \citealt{2012SoPh..275..207S,2012SoPh..275..229S}) aboard the Solar Dynamics Observatory (SDO; \citealt{2012SoPh..275....3P}), from May 2010 to February 2020. Throughout its operation, HMI keeps highly stable performance \citep[see][for the long-term stability and calibration]{2018SoPh..293...45H}.

For each day, we obtained the synoptic line-of-sight magnetograms at 0, 6, 12, and 18 UT which were rebinned from the original $4096\times 4096$ pixels to $1024\times 1024$ pixels by averaging over the values in the $4\times 4$ pixel patch.\footnote{\url{http://jsoc.stanford.edu/data/hmi/fits}} The magnetic field strength $B$ was corrected for the line-of-sight projection by dividing it by $\cos{\theta}$, the viewing angle from the disk center, and this radial field $B/\cos{\theta}$ was integrated over the disk to obtain the total radial unsigned magnetic flux of the hemisphere: $\Phi=\int|B/\cos{\theta}|\,dS$, where $dS$ is the pixel size of the magnetograms, which is constant over the disk $(=2\arcsec\times 2\arcsec)$. Although this assumption becomes less valid closer to the limb, \cite{2017SoPh..292...36L} showed that the $\cos{\theta}$ correction of $B$ in each pixel improves the estimate of $\Phi$ from the line-of-sight magnetograms. By averaging the unsigned fluxes of the four magnetograms for each day, the daily unsigned flux data were generated. We adopted the radial flux rather than the line-of-sight flux because our previous study \citep{2020ApJ...902...36T} showed that the radial flux provides stronger correlations with the irradiances. The noise level for each magnetogram was determined by fitting a Gaussian function to the distribution of the field strength \citep{2001ApJ...555..448H}.

\begin{deluxetable*}{lccccccl}
\tabletypesize{\scriptsize}
\tablecaption{Summary of the observables\label{tab:observables}}
\tablewidth{0pt}
\tablehead{
\colhead{Feature} & \colhead{$\log{T}$ (K)} & \colhead{Wavelength ({\AA})} & \colhead{Basal} & \colhead{Minimum} & \colhead{Maximum} & \colhead{Unit} & \colhead{Source}
}
\decimalcolnumbers
\startdata
Total unsigned magnetic flux & 3.8 & 6173.3 & $1.18\times 10^{23}$ & $1.16\times 10^{23}$ & $3.35\times 10^{23}$ & Mx & SDO/HMI\\
X-rays 1--8 {\AA} & 6--7 & 1--8 & 0 & $1.00\times 10^{-9}$ & $4.81\times 10^{-5}$ & W m$^{-2}$ & GOES/XRS\\
X-rays 5.2--124 {\AA} & 6--7 & 5.2--124 & $2.11\times 10^{-4}$ & $1.85\times 10^{-4}$ & $1.01\times 10^{-3}$ & W m$^{-2}$ & SORCE/XPS\\
Fe XV 284 {\AA} & 6.4 & $284.15\pm 1.50$ & $9.36\times 10^{-6}$ & $5.68\times 10^{-6}$ & $1.27\times 10^{-4}$ & W m$^{-2}$ & SORCE/XPS\\
Fe XIV 211 {\AA} & 6.3 & $211.32\pm 1.50$ & $1.20\times 10^{-5}$ & $9.88\times 10^{-6}$ & $6.75\times 10^{-5}$ & W m$^{-2}$ & SORCE/XPS\\
Fe XII 193$+$195 {\AA} & 6.2 & $193.50\pm 2.50$ & $6.16\times 10^{-5}$ & $5.66\times 10^{-5}$ & $1.72\times 10^{-4}$ & W m$^{-2}$ & SORCE/XPS\\
F10.7cm radio & $\sim$6 & $10.7\times 10^{8}$ & $68.83$ & $63.67$ & $466.57$ & sfu & DRAO\\
He II 256 {\AA}$+$blends & 4.9 & $256.30\pm 3.00$ & $5.53\times 10^{-5}$ & $5.20\times 10^{-5}$ & $1.21\times 10^{-4}$ & W m$^{-2}$ & SORCE/XPS\\
Si IV 1393 {\AA} & 4.9 & $1393.85\pm 1.30$ & $4.45\times 10^{-5}$ & $4.27\times 10^{-5}$ & $7.66\times 10^{-5}$ & W m$^{-2}$ & SORCE/SOLSTICE\\
Si IV 1402 {\AA} & 4.9 & $1402.85\pm 0.85$ & $2.32\times 10^{-5}$ & $2.25\times 10^{-5}$ & $3.91\times 10^{-5}$ & W m$^{-2}$ & SORCE/SOLSTICE\\
C II 1335 {\AA} & 4.3 & $1335.25\pm 1.90$ & $1.57\times 10^{-4}$ & $1.52\times 10^{-4}$ & $2.46\times 10^{-4}$ & W m$^{-2}$ & SORCE/SOLSTICE\\
H I 1216 {\AA} (Ly$\alpha$) & 4.3 & $1215.70\pm 2.00$ & $5.73\times 10^{-3}$ & $5.60\times 10^{-3}$ & $8.94\times 10^{-3}$ & W m$^{-2}$ & SORCE/SOLSTICE\\
Mg II k 2796 {\AA} & (3.9) & $2796.38\pm 0.78$ & $0.0136$ & $0.0135$ & $0.0180$ & W m$^{-2}$ & SORCE/SOLSTICE\\
Mg II h 2803 {\AA} & (3.9) & $2803.48\pm 0.65$ & $0.0097$ & $0.0096$ & $0.0126$ & W m$^{-2}$ & SORCE/SOLSTICE\\
Ca II K 3934 {\AA} & (3.8) & $3933.66\pm 0.50$ & $0.0114$ & $0.0111$ & $0.0130$ & W m$^{-2}$ & SORCE/SIM \& SOLIS/ISS\\
Ca II H 3968 {\AA} & (3.8) & $3968.47\pm 0.50$ & $0.0139$ & $0.0139$ & $0.0155$ & W m$^{-2}$ & SORCE/SIM \& SOLIS/ISS\\
H I 6563 {\AA} (H$\alpha$) & (3.8) & $6562.80\pm 0.50$ & $0.0369$ & $0.0360$ & $0.0448$ & W m$^{-2}$ & SORCE/SIM \& SOLIS/ISS\\
\enddata
\tablecomments{The first column shows the features, i.e., the total unsigned radial magnetic flux and the spectral lines. The second and third columns provide the formation temperature and wavelength range, respectively, for the measurement of irradiance. The temperatures are obtained from the CHIANTI database, except for the optically thick chromospheric lines, which are given in parentheses. The central wavelengths for \ion{Fe}{15} 284 {\AA} and \ion{Fe}{14} 211 {\AA} and central wavelengths and windows for \ion{Si}{4} 1393 {\AA}, \ion{Si}{4} 1402 {\AA}, \ion{C}{2} 1335 {\AA}, \ion{H}{1} 1216 {\AA} (Ly$\alpha$), \ion{Mg}{2} k 2796 {\AA}, and \ion{Mg}{2} h 2803 {\AA} are adopted from \citet{2021ApJ...908..205A}. Columns 4, 5, 6, and 7 show the basal flux, minimum and maximum values, and their physical units. $1\ {\rm sfu}= 10^{-22}\ {\rm W\ m}^{-2}\ {\rm Hz}^{-1}$. Column 8 provides the data source.}
\end{deluxetable*}

\subsection{Spectral Irradiances of the Sun}

To elucidate the atmospheric responses to the variation of magnetic flux, we analyzed the spectral irradiance data in X-ray, UV, visible, and radio bands, highlighting the atmospheric layers of the corresponding temperatures from $\log{T}=4$ to 7. The spectral lines analyzed in this study, formation temperatures, and data sources are summarized in Table \ref{tab:observables}. All obtained fluxes were converted to energy per unit area and per unit time at 1 AU from the Sun.

For the soft X-ray data, the 1--8 {\AA} band data obtained by the X-ray Sensor (XRS) on board the Geostationary Operational Environmental Satellite (GOES) were used. The ``science quality'' level 2 data from GOES-15 satellite (daily average) for the period of May 2010 to February 2020 were obtained to generate the 1--8 {\AA} daily X-ray light curve.\footnote{\url{https://www.ngdc.noaa.gov/stp/satellite/goes-r.html}} For the stability and calibration, see the User's Guide available at the corresponding webpage. Also, we referred to \citet{2015SoPh..290.3625S} for the noise level estimation ($\lesssim 3\times 10^{-9}\ {\rm W\ m^{-2}}$ at $10^{-5}\ {\rm W\ m^{-2}}$ or less).

To measure the X-ray and Extreme UV fluxes, we used the spectral irradiance data obtained by the XUV Photometer System (XPS; \citealt{2005SoPh..230..375W}) and the Solar Stellar Irradiance Comparison Experiment (SOLSTICE; \citealt{2005SoPh..230..225M}) on board the Solar Radiation and Climate Experiment (SORCE) satellite from May 2010 through February 2020.\footnote{\url{https://lasp.colorado.edu/home/sorce/data/}} Since its launch in 2003, the spacecraft and the instruments were overall stable and well calibrated, but the degradation of the battery capacity caused some observation gaps \citep{2021SoPh..296..127W}. The daily spectral irradiance data of SORCE/XPS (level 4, version 12), which covers the wavelengths from 1 to 400 {\AA} with a spectral resolution of 1 {\AA}, were used to measure the fluxes of X-ray 5.2--124 {\AA} (the ROSAT heritage band; \citealt{1982AdSpR...2d.241T,2007LRSP....4....3G}), \ion{Fe}{15} 284 {\AA}, \ion{Fe}{14} 211 {\AA}, \ion{Fe}{12} 193$+$195 {\AA} (combined), and \ion{He}{2} 256 {\AA}. Note that the spectral models in the CHIANTI atomic database were used to construct the XPS level 4 spectra. Therefore, we limited the target spectra to the representative spectra listed above to minimize the possibility that the scaling relationships between the magnetic flux and irradiances are affected by the spectral models adopted in CHIANTI. For each spectral line, the noise level was estimated using the irradiance uncertainty provided in the dataset. The central wavelength and wavelength range used to measure the irradiances are shown in Table \ref{tab:observables}. For the fluxes of \ion{Si}{4} 1393 {\AA}, \ion{Si}{4} 1402 {\AA}, \ion{C}{2} 1335 {\AA} (combination of 1334.5 and 1335.7 {\AA}), \ion{H}{1} 1216 {\AA} (Ly$\alpha$), \ion{Mg}{2} k 2796 {\AA}, and \ion{Mg}{2} h 2803 {\AA}, the daily spectral irradiance data of SORCE/SOLSITCE (level 3, version 18), covering the wavelengths from 1150 to 3100 {\AA} with a spectral resolution of 1 {\AA}, was used. In this version, the geocoronal correction is applied to the Ly$\alpha$ data. The central wavelength and spectral windows are similarly presented in Table \ref{tab:observables}.

Chromospheric lines in the visible wavelengths ($\log{T}\sim 4$) are used to diagnose the magnetic activity of the Sun and stars \citep[e.g.,][]{1972ApJ...171..565S,1978ApJ...226..379W,1995ApJ...438..269B}. The spectral data of \ion{Ca}{2} K 3934 {\AA}, \ion{Ca}{2} H 3968 {\AA}, and \ion{H}{1} 6563 {\AA} (H$\alpha$) acquired by the Integrated Sunlight Spectrometer (ISS; \citealt{2011SoPh..272..229B}) of the Synoptic Optical Long-term Investigations of the Sun (SOLIS) have been used to diagnose the chromospheric activities \citep[e.g.,][]{2007ApJ...657.1137L}. Because these spectra are provided as relative intensities with regard to the nearby continuum levels, the daily spectral irradiance data of the SORCE's Spectral Irradiance Monitor (SIM; \citealt{2005SoPh..230..141H}) (level 3, version 27), covering the spectral range from 2400 to 24200 {\AA} with a 10--340 {\AA} spectral resolution, were used to determine the absolute intensities. The spectral windows are shown in Table \ref{tab:observables}.

Note that in Table \ref{tab:observables}, the lines with different formation mechanisms are listed. For instance, the \ion{Si}{4} lines are an emission line with a single peak; the \ion{Mg}{2} and \ion{Ca}{2} lines are a self-reversal (i.e., double-peaked) line; and the H$\alpha$ line is usually in absorption on the solar disk \citep[e.g.,][]{2012ApJ...749..136L,2013ApJ...772...90L}. However, the formation of optically thick chromospheric lines is in general highly complicated due to radiative transfer processes such as multiple scattering, which may reflect different temperature regions. Although the spectral window is selected to extract the core of each line, it is difficult to narrow down the formation temperature to a single value. Therefore, the temperatures in Table \ref{tab:observables}, given in parentheses, should be considered as a reference. Also, H$\alpha$, usually in absorption on the disk, is included in the analysis to examine if this well-observed line shows any correlations with other proxies.

For the Sun’s radio emission, the F10.7 cm flux data \citep{2013SpWea..11..394T} provided by the Dominion Radio Astrophysical Observatory (DRAO) was used.\footnote{\url{https://www.spaceweather.gc.ca/forecast-prevision/solar-solaire/solarflux/sx-en.php}} We used the ``adjusted'' daily flux which is measured at local noon and corrected for the modulation of Sun–Earth distance, given in solar flux units ($1\ {\rm sfu}=10^{-22}\ {\rm W\ m}^{-2}\ {\rm Hz}^{-1}$). When no strong flare occurs, the radio flux is composed of steady thermal bremsstrahlung from the chromosphere ($\log{T}\sim 4$), gyroresonance emission from a strong magnetic field above active regions, and coronal bremsstrahlung \citep{1994ApJ...420..903G}. Because we focused on the variation component of radio flux (see below), which likely originates from the active region corona, we assumed that the formation temperature of the F10.7 cm radio flux is $\log{T}\sim 6$. Although the uncertainty is not provided in the F10.7 cm data, \citet{1994SoPh..150..305T} suggested that the average error in the flux determinations is not more than 0.5\%, and we used this value for estimating the noise levels in this study.

\subsection{Derivation of Power-law Indices}

For each of the daily total magnetic flux and irradiance curves, the basal flux was determined as the median of data points for a total of 86 days which are (1) the final one year (i.e., from March 2019 to February 2020), (2) when the daily sunspot number was zero, and (3) when the daily total magnetic flux was less than the fifth percentile of the entire period. For the 1--8 {\AA} X-ray light curve, the basal flux was set as $0\ {\rm W\ m}^{-2}$. In addition, because the SOLIS/ISS observation terminated in October 2017, basal fluxes for \ion{Ca}{2} K, \ion{Ca}{2} H, and H$\alpha$ were derived as the median values of data points of 268 days which is (1) the one-year centered in December 2008 (i.e., the activity minimum between solar cycles 23 and 24), and (2) when the daily sunspot number was zero. The corresponding basal fluxes and minimum and maximum values are shown in Table \ref{tab:observables}.

We determined the basal flux of each curve by the above method instead of simply choosing the minimum value because it is unknown whether such a minimum is truly the lowest value because of the observation gaps. To estimate the effect of differences in the basal flux computation method, we tested the three cases: (1) the selection period was the final six months; (2) the spot number was less than 10; and (3) the total magnetic flux was less than the 10th percentile. We found that the changes in the basal fluxes are typically much less than 1\% and only up to 4.2\%.

By removing the basal flux from each measurement, the excess irradiance flux $F$ and the excess total unsigned magnetic flux $\Phi$ were obtained. Thereafter, a set of double logarithmic scatter plots of a given irradiance versus total magnetic flux was developed. For each diagram, we evaluated the proportionality by fitting the data with a power-law function, $\log{F}=\alpha \log{\Phi}+\beta$, or $F\propto\Phi^{\alpha}$. We also measured the linear Pearson correlation coefficient, $CC(\log{\Phi},\log{F})$, to estimate the degree of dispersion of the data points.

In addition, we examined the dependence of power-law index $\alpha$ on different phases of the solar activity cycle. We divided the 10-year dataset into four subsets: Subset I represents the rising phase from May 2010 to August 2012; Subset I\hspace{-.1em}I the solar maximum from September 2012 to July 2015; Subset I\hspace{-.1em}I\hspace{-.1em}I the declining phase from August 2015 to November 2017; and Subset I\hspace{-.1em}V the solar minimum from December 2017 to February 2020. We evaluated the proportionality by fitting the data points in each period with a power-law function. Note that Subset I\hspace{-.1em}I is longer than the other periods because it covers the observation gap of SORCE from August 2013 to February 2014.

\subsection{Stellar Data in Literature}

The magnetic flux and luminosity data in the literature (mainly G-dwarfs with ages from 50 Myr to 4.5 Gyr) were analyzed to investigate whether the scaling laws obtained from the solar data can be applied to the sun-like stars. The total unsigned magnetic fluxes of the stars were calculated by \citet{2020AandA...635A.142K} based on magnetic observations of the spectral lines. This method measures the Zeeman broadening of atomic lines with different magnetic sensitivities to provide a better estimation of the unsigned magnetic flux which can be underestimated by the Zeeman Doppler imaging technique because of the cancellation of opposite field polarities. We used the stellar radius of each star to calculate the total unsigned magnetic flux of the visible hemisphere (see Table \ref{tab:stellar} in Appendix \ref{sec:appendix}).

In addition, from the published data of stellar luminosities in literature, irradiances of X-ray 5.2--124 {\AA}, \ion{Fe}{15} 284 {\AA}, \ion{C}{2} 1335 {\AA}, Ly$\alpha$, and \ion{Mg}{2} k$+$h (combined) at a distance of 1 AU from the stars were calculated. Stellar EUV emissions of wavelengths longer than $\sim$360 {\AA} are subject to strong absorption by the interstellar medium, which is compensated for the Ly$\alpha$ data used in this study \citep{2005ApJS..159..118W,2005ApJ...622..680R}. Table \ref{tab:stellar} shows the corresponding magnetic fluxes and irradiances. Here, we should mention that these solar-like stars have magnetically active atmospheres with starspots and associated active regions that produce highly variable XUV fluxes at short time scales (days to weeks). This does not allow the derivation of the basal flux in the manner that is defined for the Sun. Therefore, in the scatter plots (see Section \ref{sec:results}), we showed the observed magnetic fluxes and irradiances rather than the basal-flux-subtracted residuals.

\section{Results}\label{sec:results}

\begin{figure*}
\begin{center}
\includegraphics[width=0.85\textwidth]{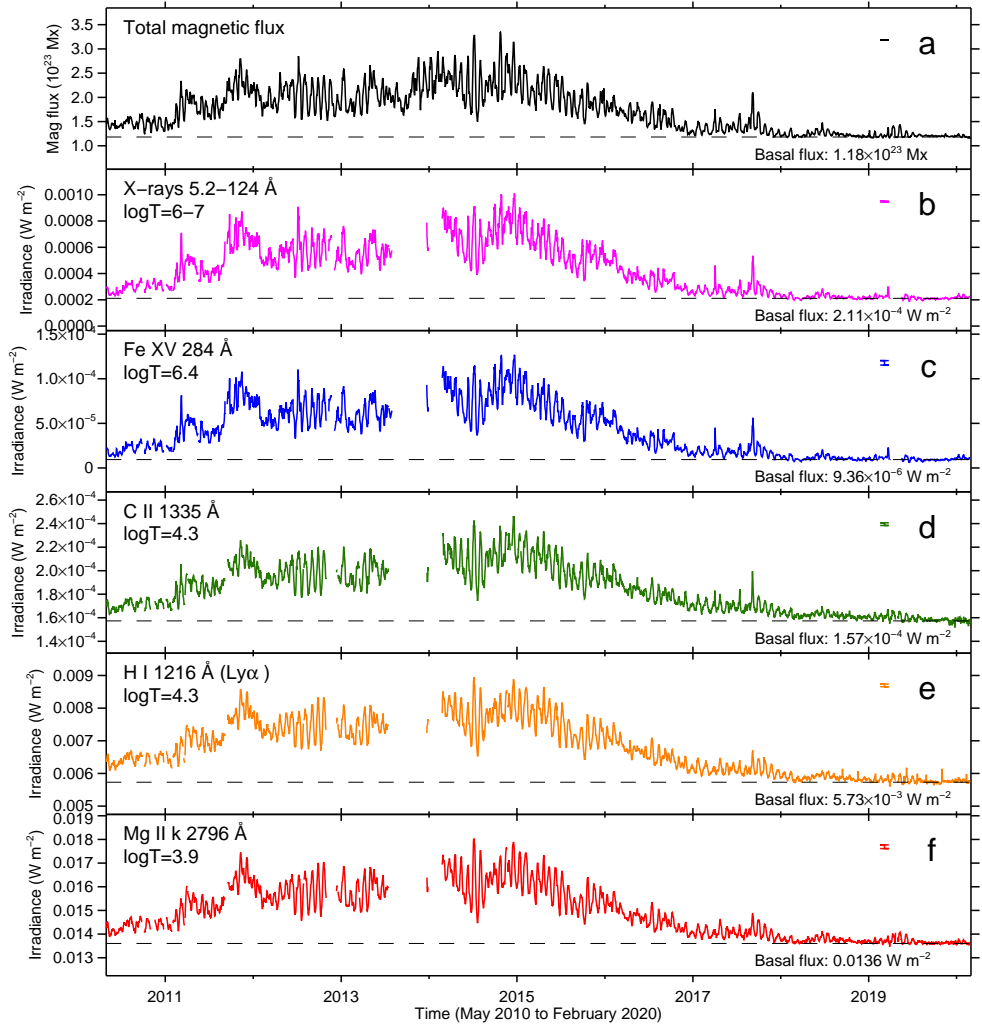}
\end{center}
\caption{Time series of total unsigned magnetic flux and spectral lines of the Sun. (a) Daily variation of total unsigned radial magnetic flux in the visible hemisphere calculated from SDO/HMI full-disk magnetograms, with a typical noise and basal flux annotated. (b)--(f) Daily variations of X-rays 5.2--124 {\AA} (the ROSAT heritage band), \ion{Fe}{15} 284 {\AA}, \ion{C}{2} 1335 {\AA}, \ion{H}{1} 1216 {\AA} (Ly$\alpha$), and \ion{Mg}{2} k 2796 {\AA}, measured by XPS and SOLSTICE on board SORCE. Corresponding formation temperatures, noise levels, and basal fluxes are annotated. All irradiance data are converted to energy per unit area and per unit time at 1 AU from the Sun. The gaps in panels (b)--(f) including the largest one from August 2013 to February 2014 are due to gaps in SORCE observations \citep{2021SoPh..296..127W}.} For each plot, the basal flux is determined as the median of data points for 86 unspotted days in the solar minimum from March 2019 to February 2020.\label{fig:lc_paper1}
\end{figure*}

The resultant time series of daily total magnetic flux and the light curves in the 5.2--124 {\AA} band and \ion{Fe}{15} 284 {\AA}, \ion{Fe}{12} 195 {\AA}, \ion{C}{2} 1335 {\AA}, and \ion{Mg}{2} k 2796 {\AA} spectral lines are shown in Figure \ref{fig:lc_paper1} (see Figure \ref{fig:lc_all} in Appendix \ref{sec:appendix} for all spectral lines). These plots demonstrate the long-term variation of solar magnetic activity peaked in 2014 with a minimum around 2019. Each spike corresponds to the transit of active regions over the solar disk. The striking correspondence between the spikes of magnetic flux and those of irradiances demonstrates that different layers of solar atmosphere heat up in response to the appearance of active regions.

\begin{figure*}
\begin{center}
\includegraphics[width=\textwidth]{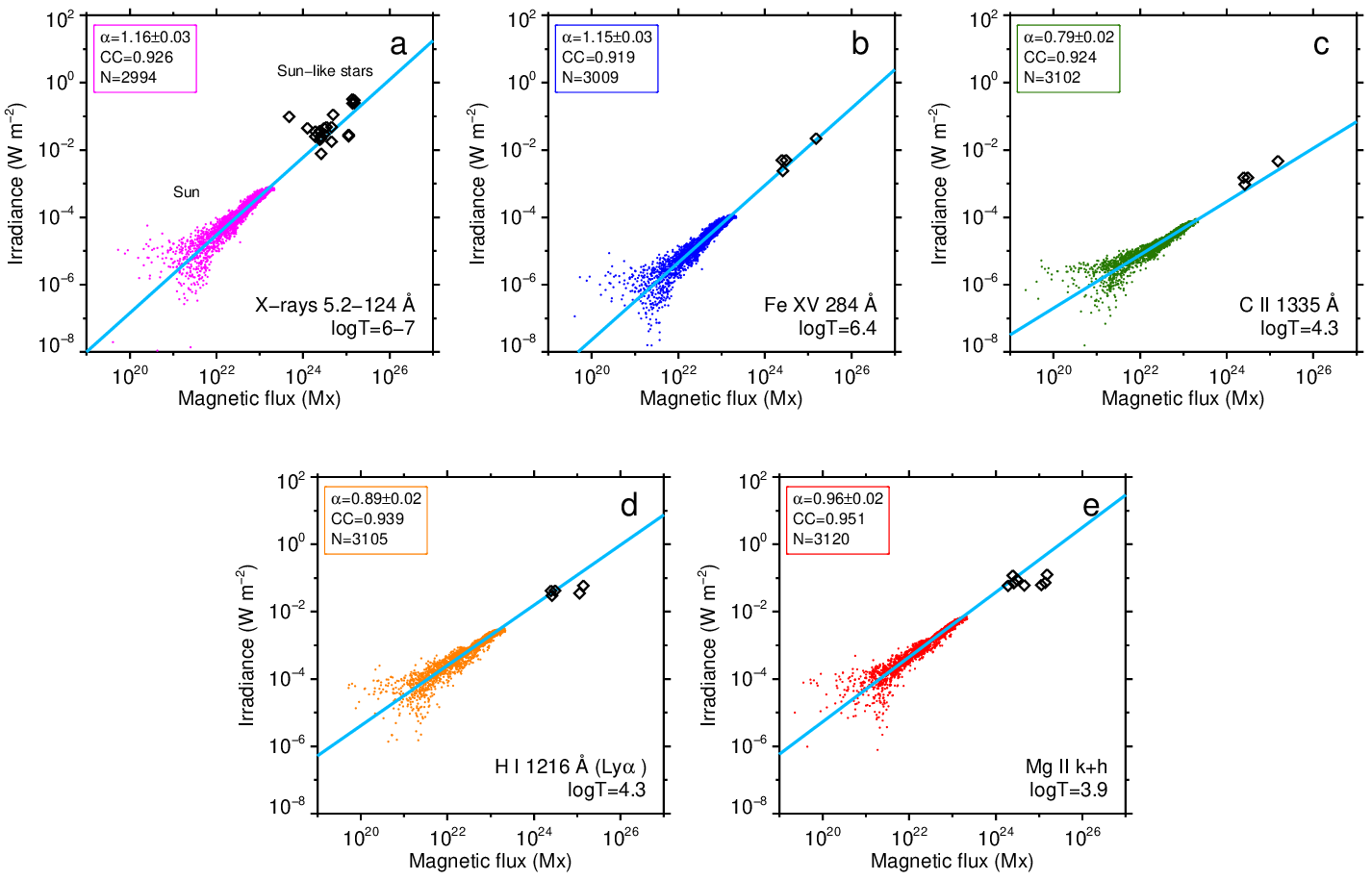}
\end{center}
\caption{Comparison of solar and stellar data. (a)--(e) Double logarithmic scatter plots (colored dots) of irradiances versus total unsigned magnetic flux for 5.2--124 {\AA}, \ion{Fe}{15} 284 {\AA}, \ion{C}{2} 1335 {\AA}, Ly$\alpha$, and \ion{Mg}{2} k$+$h (combined). Both parameters are the basal-flux-subtracted residuals and only the positive values are plotted. In each panel, a straight line shows the result of a linear fitting to the data points ($F\propto\Phi^{\alpha}$), with power-law index $\alpha$, correlation coefficient $CC$, and number of data points $N$ summarized in a box in the upper left. Diamonds are the sun-like star data in literature.\label{fig:cc_paper1}}
\end{figure*}

Each panel of Figure \ref{fig:cc_paper1} shows a scatter plot of the variation of a given solar irradiance from its background level (basal flux) versus that of the total unsigned magnetic flux (see Figure \ref{fig:cc_all} in Appendix \ref{sec:appendix} for all spectral lines). In each panel, only the data points where both the irradiance $F$ and unsigned flux $\Phi$ are positive are plotted, and the number of such data points, $N$, is shown in the small box in the upper left corner. The power-law index $\alpha$ for the linear fit ($\log{F}=\alpha\log{\Phi}+\beta$) and the correlation coefficient $CC(\log{\Phi},\log{F})$ are also provided. The power-law index $\alpha$, offset $\beta$, correlation coefficient $CC$, and number of data points $N$ for all scatter plots are summarized in Table \ref{tab:powerlaw}.

Here, one may find that the data points are often dispersed and sometimes show a two-branched shape at the lower end (the bottom left corner of each panel). Considering the results of, e.g., \citet{2003ApJ...598.1387P}, in which the data points are located on the power-law line even in the range of $\Phi=10^{16}\ {\rm Mx}$, the dispersed or branched tails in this study probably reflect the measurement noise rather than physical origins.

\begin{deluxetable*}{lcccccc}
\tablecaption{Power-law indices and correlations between irradiance and total magnetic flux\label{tab:powerlaw}}
\tablewidth{0pt}
\tablehead{
\colhead{Feature} & \colhead{$\log{T}$ (K)} & \colhead{Power-law index $\alpha$} & \colhead{Offset $\beta$} & \colhead{Correlation coefficient $CC$} & \colhead{Data points $N$} & \colhead{LS deviation}
}
\decimalcolnumbers
\startdata
X-rays 1--8 {\AA} & 6--7 & $1.42\pm 0.04$ & $-38.6\pm 0.8$ & 0.893 & 3243 & 0.431\\
X-rays 5.2--124 {\AA} & 6--7 & $1.16\pm 0.03$ & $-29.9\pm 0.7$ & 0.926 & 2994 & 0.247\\
Fe XV 284 {\AA} & 6.4 & $1.15\pm 0.03$ & $-30.6\pm 0.7$ & 0.919 & 3009 & 0.258\\
Fe XIV 211 {\AA} & 6.3 & $1.15\pm 0.03$ & $-30.9\pm 0.7$ & 0.924 & 2998 & 0.248\\
Fe XII 193$+$195 {\AA} & 6.2 & $1.14\pm 0.03$ & $-30.5\pm 0.7$ & 0.925 & 2998 & 0.246\\
F10.7cm radio & $\sim$6 & $1.24\pm 0.03$ & $-26.8\pm 0.7$ & 0.939 & 3200 & 0.225\\
He II 256 {\AA} & 4.9 & $1.14\pm 0.03$ & $-30.8\pm 0.7$ & 0.923 & 3001 & 0.249\\
Si IV 1393 {\AA} & 4.9 & $0.90\pm 0.02$ & $-25.3\pm 0.5$ & 0.923 & 3089 & 0.215\\
Si IV 1402 {\AA} & 4.9 & $0.83\pm 0.02$ & $-24.1\pm 0.5$ & 0.914 & 3096 & 0.214\\
C II 1335 {\AA} & 4.3 & $0.79\pm 0.02$ & $-22.5\pm 0.5$ & 0.924 & 3102 & 0.193\\
H I 1216 {\AA} (Ly$\alpha$) & 4.3 & $0.89\pm 0.02$ & $-23.3\pm 0.5$ & 0.939 & 3105 & 0.193\\
Mg II k 2796 {\AA} & (3.9) & $0.95\pm 0.02$ & $-24.4\pm 0.5$ & 0.949 & 3120 & 0.187\\
Mg II h 2803 {\AA} & (3.9) & $0.97\pm 0.03$ & $-25.2\pm 0.6$ & 0.944 & 3097 & 0.200\\
Mg II k$+$h & (3.9) & $0.96\pm 0.02$ & $-24.5\pm 0.6$ & 0.951 & 3120 & 0.187\\
Ca II K 3934 {\AA} & (3.8) & $0.87\pm 0.03$ & $-23.1\pm 0.8$ & 0.723 & 1755 & 0.214\\
Ca II H 3968 {\AA} & (3.8) & $0.86\pm 0.04$ & $-22.7\pm 0.9$ & 0.539 & 1624 & 0.273\\
H I 6563 {\AA} (H$\alpha$) & (3.8) & $-1.46\pm 0.14$ & $ 29.9\pm 3.1$ & $-0.152$ & 1487 & 0.643\\
\enddata
\tablecomments{The first and second columns show the spectral lines and their formation temperatures, respectively. Columns 3, 4, 5, and 6 provide the power-law index $\alpha$, offset $\beta$, correlation coefficient $CC$, and the number of data points $N$ of each double logarithmic scatter plot of irradiance versus total magnetic flux. Column 7 presents the least-square deviation of the linear fit to the double logarithmic plot.}
\end{deluxetable*}

In Figure \ref{fig:cc_paper1}, the total magnetic fluxes and irradiances of the solar-type stars that have been measured in the past are overplotted. The stellar data are located on the extensions of the power-law relations of the solar data, indicating that the atmospheric responses to the magnetic flux are universal for the Sun and sun-like stars, regardless of age or activity.

\begin{figure*}
\begin{center}
\includegraphics[width=0.7\textwidth]{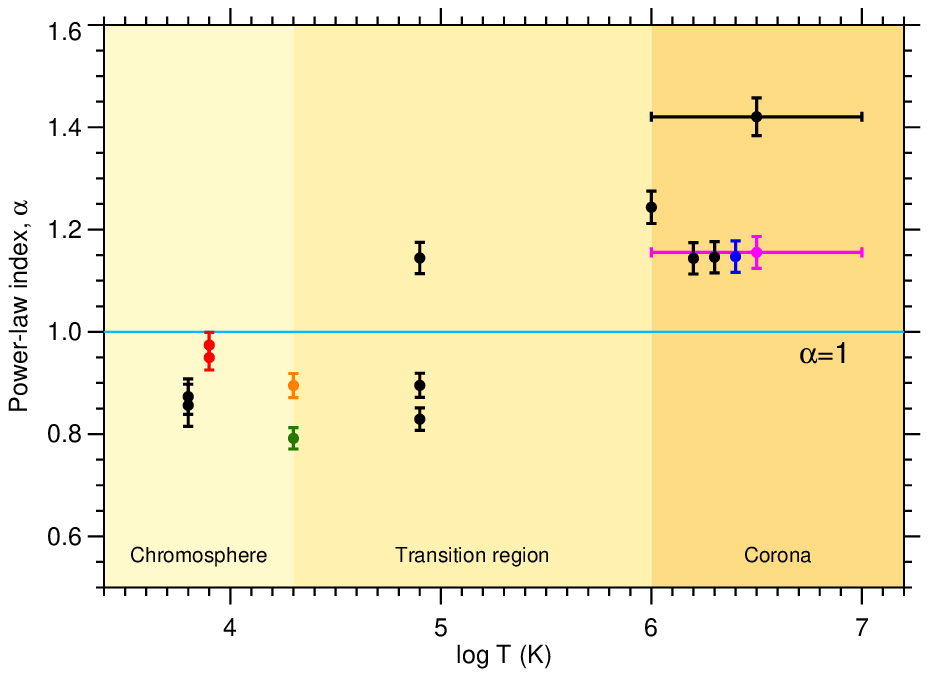}
\end{center}
\caption{Temperature dependence of the power-law indices. Power-law indices $\alpha$ for the solar data, obtained by fitting linear functions to double logarithmic scatter plots of total unsigned magnetic flux and irradiances of various spectral lines, are plotted as a function of temperature. Errors on $\alpha$ are indicated by vertical bars, while the horizontal bars show the temperature ranges for two X-ray data, 1--8 {\AA} and 5.2--124 {\AA}. Color symbols denote those compared with the stellar data: 5.2--124 {\AA} (purple), \ion{Fe}{15} 284 {\AA} (blue), \ion{C}{2} 1335 {\AA} (green), Ly$\alpha$ (orange), and \ion{Mg}{2} k and h (red). H$\alpha$ is omitted here because it exhibited a negative proportionality with magnetic flux (i.e., $\alpha<0$).\label{fig:pl_paper1}}
\end{figure*}

Figure \ref{fig:pl_paper1} shows the power-law indices, $\alpha$, as a function of formation temperature. The X-ray fluxes ($\log{T}=6$--7) are known to show a power-law relation with an exponent of slightly above unity ($\alpha=1.1$--1.8) for the Sun and sun-like stars (Section \ref{sec:introduction}). We confirmed this trend and also revealed that other coronal XUV lines and radio fluxes show much the same relationship. On the other hand, for the temperatures characteristic of the transition-region and chromospheric range ($\log{T}<6$), the power-law index decreases below unity, indicating that the atmospheric response to the surface magnetic flux falls below linearity.

Note here that H$\alpha$ is omitted from Figure \ref{fig:pl_paper1} because it has an inverse proportionality with magnetic flux ($\alpha<0$). The light curve of H$\alpha$ enhanced only in the declining phase of the solar cycle (bottom panel of Figure \ref{fig:lc_all}), and as a result, it showed a negative correlation against magnetic flux with a large scatter ($CC=-0.152$: Figure \ref{fig:cc_all}). A recent Sun-as-a-star observation by \citet{2019A&A...627A.118M} indicates a similar anti-correlation between \ion{Ca}{2} K and H$\alpha$, and this may be attributed to the difference in the line formation processes: while the \ion{Ca}{2} K flux increases monotonically with increasing chromospheric temperature or mass, H$\alpha$ can be absorption or emission depending on the temperature or mass \citep{1987ApJ...323..316C,1989A&A...219..239R}.

\begin{figure*}
\begin{center}
\includegraphics[width=0.7\textwidth]{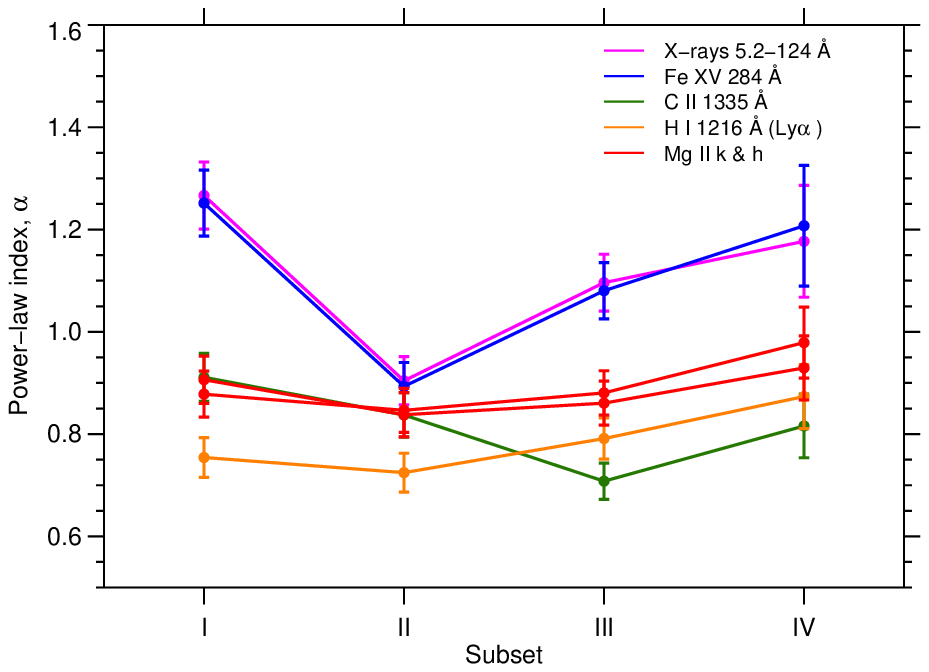}
\end{center}
\caption{Power-law indices $\alpha$ of the representative lines that are measured for the data points in different subsets of the solar activity cycle, which are (I) the rising phase from May 2010 to August 2012; (I\hspace{-.1em}I) the solar maximum from September 2012 to July 2015; (I\hspace{-.1em}I\hspace{-.1em}I) the declining phase from August 2015 to November 2017; and (I\hspace{-.1em}V) the solar minimum from December 2017 to February 2020. Subset II has a longer time interval than the other periods because it covers the observation gap of SORCE from August 2013 to February 2014.\label{fig:pl_phase_paper}}
\end{figure*}

The phase dependence of $\alpha$ is summarized in Figure \ref{fig:pl_phase_paper}, which shows the $\alpha$ values of representative spectral lines that are measured in four different phases of the solar activity cycle (see Table \ref{tab:powerlaw2} for all proxies). It is clearly seen that $\alpha$ takes its minimum in Subset I\hspace{-.1em}I or I\hspace{-.1em}I\hspace{-.1em}I, i.e., at the solar maximum or in the declining phase, while it becomes larger in Subset I or I\hspace{-.1em}V, i.e., in the rising or minimum phase. In Table \ref{tab:powerlaw2}, the trends are not clear for the chromospheric lines (\ion{Ca}{2} K, \ion{Ca}{2} H, and H$\alpha$), possibly because of their narrow dynamic ranges compared to the other proxies.

\section{Discussion and Conclusion}\label{sec:discussion}

The temperature dependence of power-law scaling relationships between the magnetic flux and radiation fluxes discovered in this study is a powerful tool for understanding the heating mechanism of stellar atmospheres that is common to the Sun and sun-like stars.

For the coronal regime ($\log{T}>6$), where the power-law index exceeds unity, theoretical explanations have been provided \citep{1998ApJ...508..885F,2020A&A...640A.119Z,2020ApJ...901...70T}. These include the heating via Alfv\'{e}n waves \citep{2011ApJ...736....3V} and electric current sheets \citep{1972ApJ...174..499P,1988ApJ...330..474P}. In the first mechanism, the Alfv\'{e}n waves excited by shuffling convective motions in the solar/stellar surface propagate upwards along the magnetic field lines and dissipate energy into heat. In the second model, the energy is dissipated in current sheets within braided field-line structures driven by photospheric convection as a nanoscale version of solar flares (nanoflares). Recent numerical modeling of coronal loops considering the effect of turbulent dissipation of Alfv\'{e}n waves successfully reproduced the power-law index of $\alpha=1.19$ for the Sun and sun-like stars \citep{2021A&A...656A.111S}. Young, fast-rotating stars are known to exhibit high X-ray quiescent luminosity, which is attributed to the enhanced dynamo activity driven by fast stellar rotation \citep{2003A&A...397..147P,2011ApJ...743...48W}. Our result of power-law indices being $\alpha>1$ for $\log{T}>6$ provides the robust view that the same heating mechanism---the energy supply by surface convection, transportation along the magnetic field, and dissipation in the atmosphere---is commonly at work in these stars, regardless of age or activity, to generate the hot coronae.

The \ion{Ca}{2} K emission ($\log{T}\sim 3.8$) has been found to correlate with surface magnetic flux with power-law indices $\alpha=0.6$--1 \citep[e.g.,][]{1975ApJ...200..747S,1989ApJ...337..964S}, and our study reveals that the tendency of $\alpha\lesssim 1$ is true for the chromospheric lines of sun-like stars in general. \citet{1989ApJ...337..964S} has attributed this tendency to the geometrical effect of magnetic flux tubes, where vertical expansion of the flux tubes is restricted because the tubes are concentrated at the boundaries of convection cells, resulting in a small effective area of the chromosphere. The flux-tube expansion also plays a key role in reflecting the ascending Alfv\'{e}n waves via drastic density reduction \citep{2005ApJS..156..265C}, and this may suppress the efficiency of chromospheric heating by Alfv\'{e}n waves to a sublinear proportionality with photospheric magnetic flux.

The power-law indices $\alpha$ were also found to have a solar cycle dependence. Specifically, they reach minimum values during the maximum phase of the cycle. One possible explanation is that, at maximum, the solar atmosphere is filled with magnetic flux sourced from active regions that cover the sphere, and thus the atmospheric heating becomes saturated. That is, compared to the quiescent phases, the atmosphere is not effectively heated any more even if the new magnetic flux is supplied to the surface. However, the determination of what physical mechanism causes the saturation requires further assessment.

The power-law scaling relationships obtained in this study provide a means to empirically estimate the irradiances over a wide range of wavelengths from sun-like stars. This allows us to calculate the quiescent radiation fluxes, specifically of XUV emissions, to the (exo)planetary atmospheres and their surfaces derived from the magnetic flux distribution of observed or modeled sun-like planet-hosting stars. Once the total radial unsigned magnetic flux in the visible hemisphere is obtained from observations or modeling, the irradiance of a given line, $F_{1}$, can be estimated using the values in Tables \ref{tab:observables} and \ref{tab:powerlaw} as $F_{1}=\beta(\Phi_{1}-\Phi_{0})^{\alpha}+F_{0}$, where $\Phi_{1}$, $\Phi_{0}$, and $F_{0}$ are the observed and basal magnetic fluxes and basal irradiance, respectively.

This stellar output provides observationally constrained input for studying the response of (exo)planetary atmospheres in terms of associated atmospheric escape and impact on atmospheric chemistry, the critical factors of exoplanetary habitability and associated potential biosignatures. On the other hand, because most of the stellar observations are snapshots and usually not synchronized with magnetic field measurements, it is possible that only selected phases of activity variations are reflected in the analysis, causing a scatter in the stellar data of Figure \ref{fig:cc_paper1}. Long-term, multi-wavelength monitoring of irradiances and magnetic fields with future solar and space telescopes, probably leveraging numerical modeling, may further reveal the physical mechanisms behind atmospheric responses to surface magnetic field \citep{2017ApJ...834...56T,2020ApJ...902...36T} and lead to the understanding of stellar flare eruptions \citep{2012Natur.485..478M,2016ApJ...829...23D}.

\begin{acknowledgments}
The authors thank the anonymous referee for the constructive comments, which improved the quality of the paper. Data are courtesy of the science teams of SDO, GOES, SORCE, SOLIS, and DRAO. HMI is an instrument on board SDO, a mission for NASA's Living With a Star program. ISS data were acquired by SOLIS instruments operated by NISP/NSO/AURA/NSF. This work was supported by JSPS KAKENHI Grant Nos. JP20KK0072 (PI: S. Toriumi), JP21H01124 (PI: T. Yokoyama), and JP21H04492 (PI: K. Kusano), and by the NINS program for cross-disciplinary study (Grant Nos. 01321802 and 01311904) on Turbulence, Transport, and Heating Dynamics in Laboratory and Astrophysical Plasmas: ``SoLaBo-X.'' V. S. Airapetian was supported by the GSFC Sellers Exoplanet Environments Collaboration (SEEC), which is funded by the NASA Planetary Science Division's Internal Scientist Funding Model (ISFM), NASA's TESS Cycle 1, HST Cycle 27 and NICER Cycle 2 project funds.
\end{acknowledgments}

\vspace{5mm}

\appendix

\section{Additional Figures and Tables}\label{sec:appendix}

Figure \ref{fig:lc_all} provides the time series of total radial unsigned flux and irradiances of all spectral lines, while Figure \ref{fig:cc_all} shows the scatter plots of total magnetic flux and irradiances of all lines. Table \ref{tab:stellar} summarizes the characteristics, total hemispheric magnetic fluxes, and irradiances of the stars analyzed with the references. Table \ref{tab:powerlaw2} presents the power-law indices $\alpha$ of all spectral lines that are measured in four different phases of the solar activity cycle.

\begin{figure*}
\begin{center}
\includegraphics[width=0.85\textwidth]{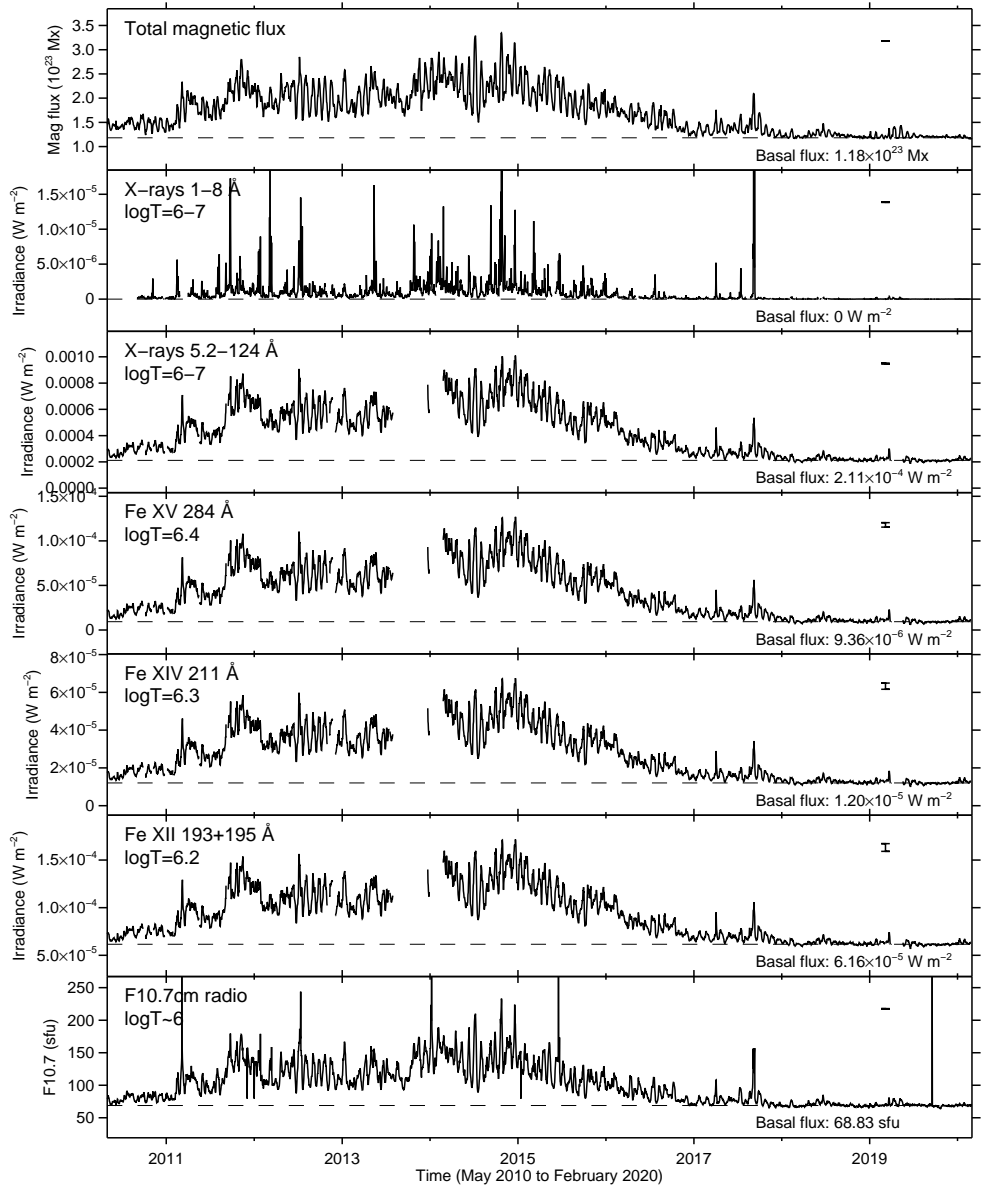}
\end{center}
\caption{Time series of total unsigned magnetic flux and irradiances of all spectral lines. In each panel, the typical noise level is shown as an error bar on the right. The basal flux is shown as a horizontal dashed line with the value provided at the bottom right.\label{fig:lc_all}}
\end{figure*}

\addtocounter{figure}{-1}
\begin{figure*}
\begin{center}
\includegraphics[width=0.85\textwidth]{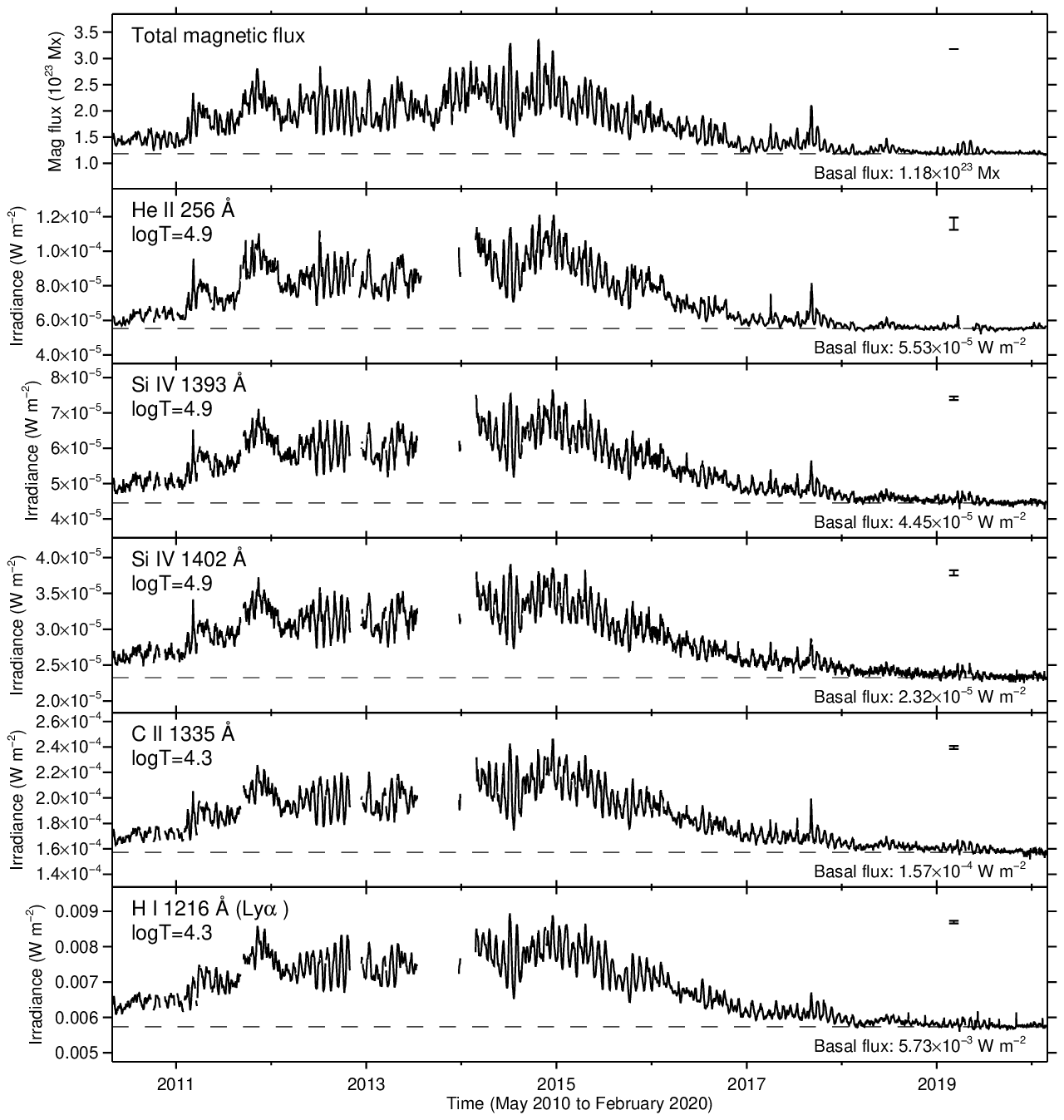}
\end{center}
\caption{{\it Continued.}}
\end{figure*}

\addtocounter{figure}{-1}
\begin{figure*}
\begin{center}
\includegraphics[width=0.85\textwidth]{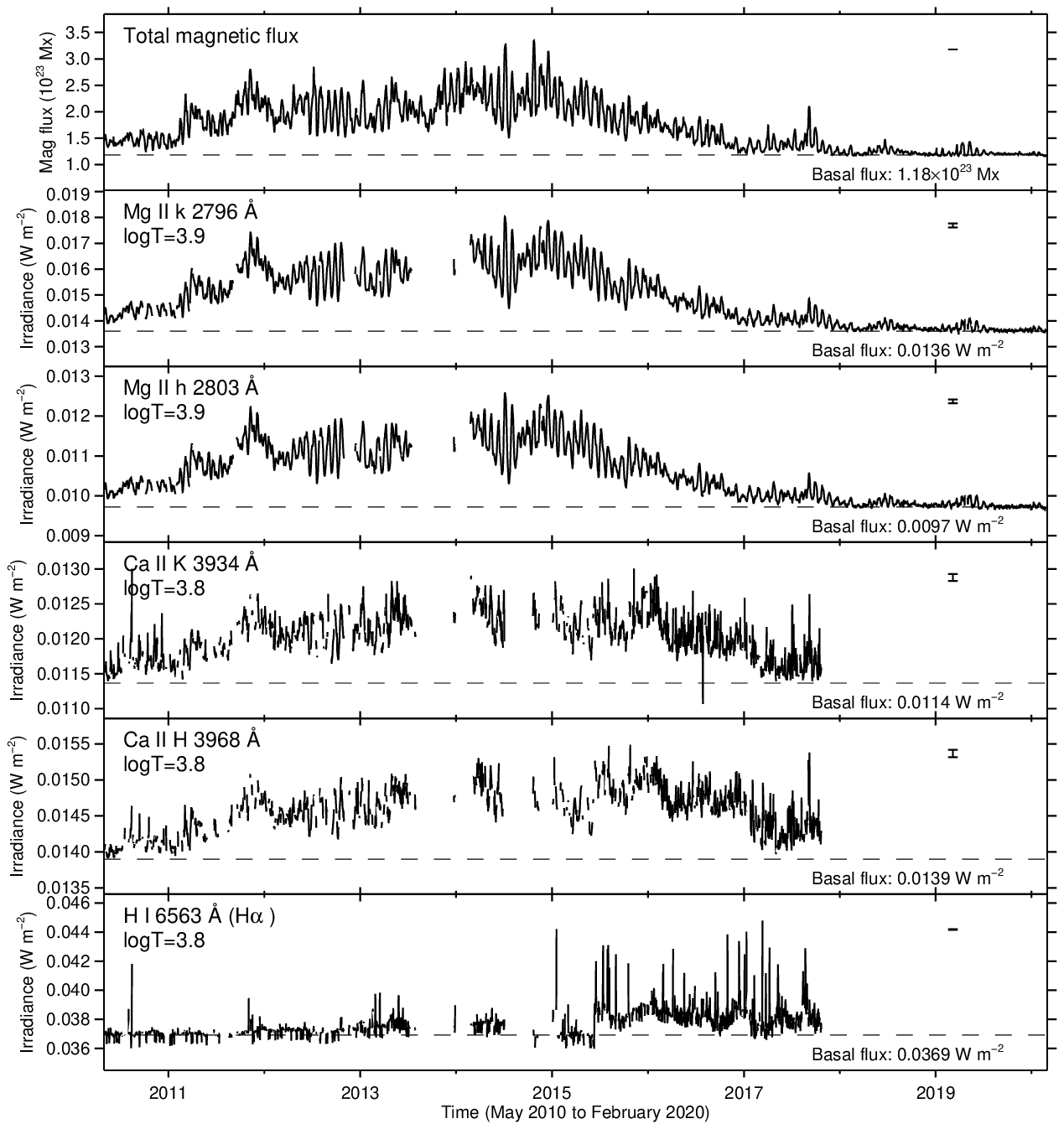}
\end{center}
\caption{{\it Continued.}}
\end{figure*}

\begin{figure*}
\begin{center}
\includegraphics[width=0.95\textwidth]{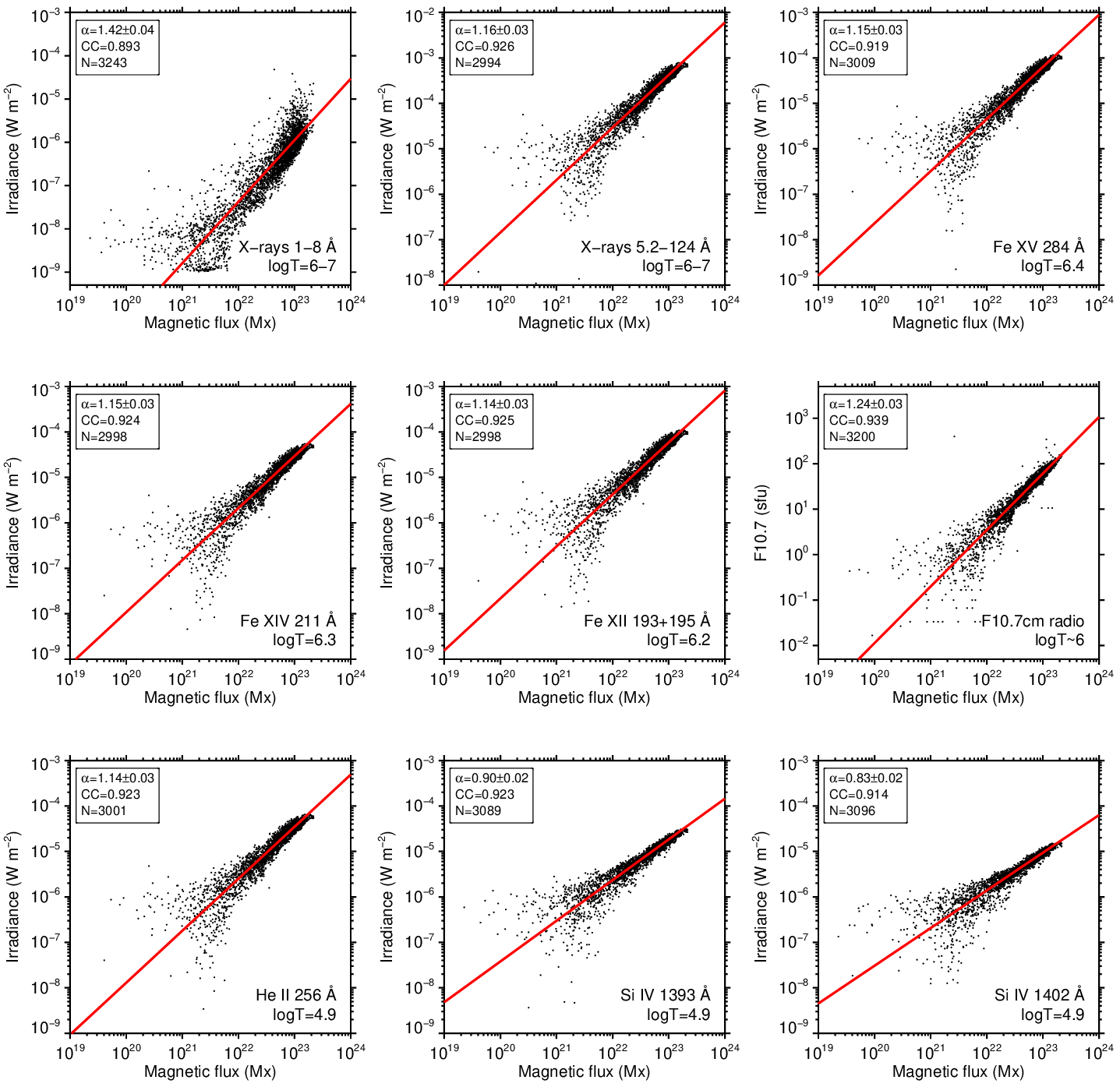}
\end{center}
\caption{Double logarithmic scatter plots of irradiance versus total magnetic flux. In each panel, the red line shows the results of a linear fitting to the double logarithmic plots. The power-law index $\alpha$, correlation coefficient $CC$, and number of data points $N$ are shown at the top left of each panel.\label{fig:cc_all}}
\end{figure*}

\addtocounter{figure}{-1}
\begin{figure*}
\begin{center}
\includegraphics[width=0.95\textwidth]{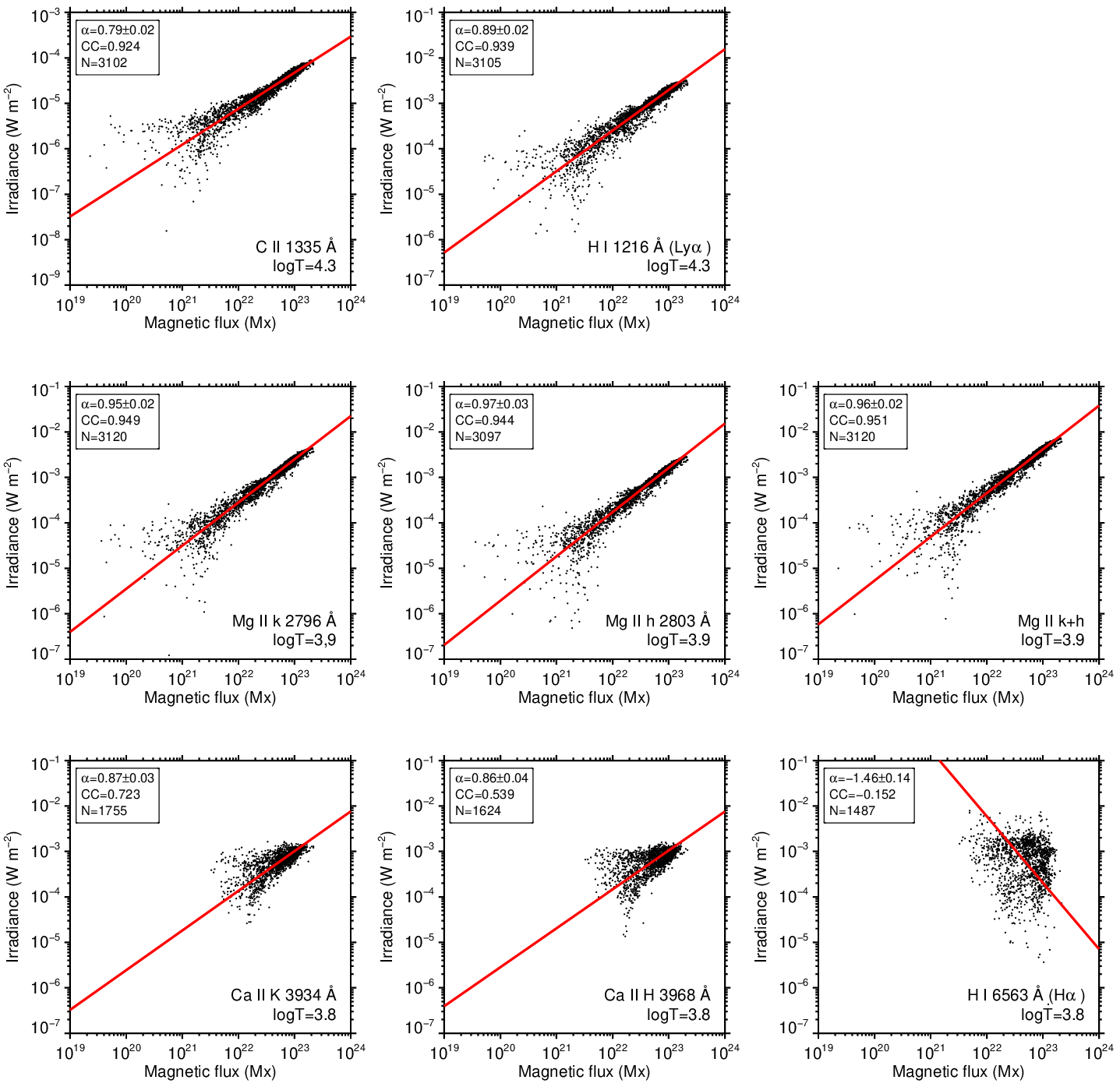}
\end{center}
\caption{{\it Continued.}}
\end{figure*}

\begin{longrotatetable}
\begin{deluxetable*}{cccccccccccccc}
\tabletypesize{\scriptsize}
\tablecaption{Characteristics of the sun-like stars\label{tab:stellar}}
\tablewidth{0pt}
\tablehead{
\colhead{HD} & \colhead{Name} & \colhead{Sp type} & \colhead{$T_{\rm eff}$} & \colhead{$\log{g}$} & \colhead{Age} & \colhead{$P_{\rm rot}$} & \colhead{$R$} & \colhead{$\Phi$} & \colhead{X-rays 5.2--124 {\AA}} & \colhead{\ion{Fe}{15} 284 {\AA}} & \colhead{\ion{C}{2} 1335 {\AA}} & \colhead{Ly$\alpha$} & \colhead{\ion{Mg}{2} k$+$h}\\
\colhead{} & \colhead{} & \colhead{} & \colhead{(K)} & \colhead{} & \colhead{(Myr)} & \colhead{(d)} & \colhead{($R_{\odot}$)} & \colhead{(Mx)} & \colhead{(W m$^{-2}$)} & \colhead{(W m$^{-2}$)} & \colhead{(W m$^{-2}$)} & \colhead{(W m$^{-2}$)} & \colhead{(W m$^{-2}$)}
}
\decimalcolnumbers
\startdata
1835 & BE Cet & G3V & 5837 & 4.47 & 600 & 7.78 & 1.00 & $4.55\times 10^{24}$ & $4.80\times 10^{-2}$, $1.78\times 10^{-2}$ & \nodata & \nodata & \nodata & $6.04\times 10^{-2}$\\
20630 & $\kappa^{1}$ Cet & G5V & 5742 & 4.49 & 600 & 9.3 & 0.95 & $2.61\times 10^{24}$ & $2.19\times 10^{-2}$, $2.56\times 10^{-2}$ & $2.40\times 10^{-3}$ & $9.50\times 10^{-4}$ & $3.01\times 10^{-2}$ & $7.09\times 10^{-2}$\\
39587 & $\chi^{1}$ Ori & G0V & 5882 & 4.34 & 500 & 4.83 & 1.05 & $2.47\times 10^{24}$ & $3.48\times 10^{-2}$, $3.73\times 10^{-2}$ & $5.00\times 10^{-3}$ & $1.52\times 10^{-3}$ & $4.16\times 10^{-2}$ & $1.18\times 10^{-1}$\\
56124 &  & G0V & 5848 & 4.46 & 4500 & 18 & 1.01 & $4.78\times 10^{23}$ & $9.79\times 10^{-2}$ & \nodata & \nodata & \nodata & \nodata\\
72905 & $\pi^{1}$ Uma & G1.5V & 5873 & 4.44 & 500 & 4.9 & 0.95 & $3.08\times 10^{24}$ & $4.48\times 10^{-2}$, $2.96\times 10^{-2}$ & $5.00\times 10^{-3}$ & $1.52\times 10^{-3}$ & $4.22\times 10^{-2}$ & $8.93\times 10^{-2}$\\
73350 & V401 Hya & G5V & 5802 & 4.48 & 510 & 12.3 & 0.98 & $2.43\times 10^{24}$ & $2.05\times 10^{-2}$ & \nodata & \nodata & \nodata & \nodata\\
76151 &  & G3V & 5790 & 4.55 & 3600 & 20.5 & 1.00 & $2.62\times 10^{24}$ & $7.78\times 10^{-3}$ & \nodata & \nodata & \nodata & \nodata\\
82558 & LQ Hya & K1V & 5000 & 4.00 & 50 & 1.601 & 0.71 & $1.39\times 10^{25}$ & $3.24\times 10^{-1}$, $2.43\times 10^{-1}$ & \nodata & \nodata & $5.91\times 10^{-2}$ & $7.27\times 10^{-2}$\\
129333 & EK Dra & G1.5V & 5845 & 4.47 & 120 & 2.606 & 0.97 & $1.52\times 10^{25}$ & $3.03\times 10^{-1}$, $2.52\times 10^{-1}$ & $2.20\times 10^{-2}$ & $4.70\times 10^{-3}$ & \nodata & $1.26\times 10^{-1}$\\
131156 & $\xi$ Boo A & G7V & 5570 & 4.65 & 200 & 6.4 & 0.83 & $1.13\times 10^{25}$ & $2.58\times 10^{-2}$, $2.83\times 10^{-2}$ & \nodata & \nodata & $3.53\times 10^{-2}$ & $6.19\times 10^{-2}$\\
166435 &  & G1IV & 5843 & 4.44 & 3800 & 3.43 & 0.99 & $4.94\times 10^{24}$ & $1.12\times 10^{-1}$ & \nodata & \nodata & \nodata & \nodata\\
175726 &  & G0V & 5998 & 4.41 & 500 & 3.92 & 1.06 & $1.26\times 10^{24}$ & $4.48\times 10^{-2}$ & \nodata & \nodata & \nodata & \nodata\\
190771 &  & G2V & 5834 & 4.44 & 2700 & 8.8 & 1.01 & $3.48\times 10^{24}$ & $4.80\times 10^{-2}$ & \nodata & \nodata & \nodata & \nodata\\
206860 & HN Peg & G0V & 5974 & 4.47 & 260 & 4.55 & 1.04 & $1.92\times 10^{24}$ & $3.56\times 10^{-2}$, $2.52\times 10^{-2}$ & \nodata & \nodata & \nodata & $5.90\times 10^{-2}$\\
\hline
Sun & (mean) & G2V & 5777 & 4.44 & 4600 & 25.4 & 1.00 & $1.73\times 10^{23}$ & $4.24\times 10^{-4}$ & $4.12\times 10^{-5}$ & $1.84\times 10^{-4}$ & $6.77\times 10^{-3}$ & $2.55\times 10^{-2}$\\
 & (median) & & & & & & & $1.67\times 10^{23}$ & $3.87\times 10^{-4}$ & $3.59\times 10^{-5}$ & $1.82\times 10^{-4}$ & $6.69\times 10^{-3}$ & $2.52\times 10^{-2}$\\
 & (max) & & & & & & & $3.35\times 10^{23}$ & $1.01\times 10^{-3}$ & $1.27\times 10^{-4}$ & $2.46\times 10^{-4}$ & $8.94\times 10^{-3}$ & $3.06\times 10^{-2}$\\
 & (min) & & & & & & & $1.16\times 10^{23}$ & $1.85\times 10^{-4}$ & $5.68\times 10^{-6}$ & $1.52\times 10^{-4}$ & $5.60\times 10^{-3}$ & $2.32\times 10^{-2}$\\
\enddata
\tablerefs{\citet{1993BICDS..43....5T}, \citet{2005ApJS..159..141V}, \citet{2012MNRAS.427..343M}, \citet{2010MNRAS.403.1368G}, \citet{2015AandA...581A..69C}, \citet{1999AandA...352..555A}, \citet{2014MNRAS.441.2361V}, \citet{2016AandA...593A..35R}, \citet{2016AandA...593A..35R}, \citet{2016AandA...590A.133O}, \citet{2019ApJ...876..118S}, \citet{2020AandA...635A.142K}, \citet{2005ApJ...622..653T}, \citet{2005ApJ...622..680R}, \citet{2007ApJS..168..297T}, \citet{2010ApJ...717.1279W}, \citet{1997ApJ...483..947G}, \citet{2005ApJS..159..118W}, \citet{1990ApJ...351..492S}, \citet{1994svsp.coll..206D}}
\tablecomments{The HD number, name, spectral type, effective temperature, surface gravity, age, rotation period, and radius of the stars are shown in Columns 1--8. Column 9 shows the total hemispheric magnetic flux estimated based on the Zeeman broadening of the spectral lines. Columns 10--14 show the irradiances of X-ray 5.2--124 {\AA}, \ion{Fe}{15} 284 {\AA}, \ion{C}{2} $1334.5+1335.7$ {\AA}, Ly$\alpha$, and \ion{Mg}{2} k$+$h (combined) in the literature, all converted to the values at 1 AU from the stars. For X-rays, multiple observations are shown (if they exist).}
\end{deluxetable*}
\end{longrotatetable}

\begin{deluxetable*}{lcccccc}
\tablecaption{Power-law indices for different phases of the solar activity cycle\label{tab:powerlaw2}}
\tablewidth{0pt}
\tablehead{
\colhead{Feature} & \colhead{$\log{T}$ (K)} & \colhead{Power-law index $\alpha$} & \colhead{Subset I $\alpha_{\rm I}$} & \colhead{Subset I\hspace{-.1em}I $\alpha_{\rm I\hspace{-.1em}I}$} & \colhead{Subset I\hspace{-.1em}I\hspace{-.1em}I $\alpha_{\rm I\hspace{-.1em}I\hspace{-.1em}I}$} & \colhead{Subset I\hspace{-.1em}V $\alpha_{\rm I\hspace{-.1em}V}$}
}
\decimalcolnumbers
\startdata
X-rays 1--8 {\AA} & 6--7 & $1.42\pm 0.04$ & $1.93\pm 0.12$ & $2.45\pm 0.13$ & $1.78\pm 0.10$ & $1.19\pm 0.09$ \\
X-rays 5.2--124 {\AA} & 6--7 & $1.16\pm 0.03$ & $1.27\pm 0.07$ & $0.90\pm 0.05$ & $1.10\pm 0.06$ & $1.18\pm 0.11$ \\
Fe XV 284 {\AA} & 6.4 & $1.15\pm 0.03$ & $1.25\pm 0.06$ & $0.89\pm 0.05$ & $1.08\pm 0.06$ & $1.21\pm 0.12$ \\
Fe XIV 211 {\AA} & 6.3 & $1.15\pm 0.03$ & $1.25\pm 0.06$ & $0.90\pm 0.05$ & $1.09\pm 0.06$ & $1.16\pm 0.11$ \\
Fe XII 193$+$195 {\AA} & 6.2 & $1.14\pm 0.03$ & $1.25\pm 0.06$ & $0.90\pm 0.05$ & $1.09\pm 0.06$ & $1.14\pm 0.11$ \\
F10.7cm radio & $\sim$6 & $1.24\pm 0.03$ & $1.40\pm 0.07$ & $1.17\pm 0.05$ & $1.39\pm 0.07$ & $1.21\pm 0.11$ \\
He II 256 {\AA} & 4.9 & $1.14\pm 0.03$ & $1.25\pm 0.06$ & $0.90\pm 0.05$ & $1.08\pm 0.06$ & $1.16\pm 0.11$ \\
Si IV 1393 {\AA} & 4.9 & $0.90\pm 0.02$ & $0.85\pm 0.04$ & $0.79\pm 0.04$ & $0.76\pm 0.04$ & $0.94\pm 0.07$ \\
Si IV 1402 {\AA} & 4.9 & $0.83\pm 0.02$ & $0.77\pm 0.04$ & $0.76\pm 0.04$ & $0.70\pm 0.04$ & $0.82\pm 0.07$ \\
C II 1335 {\AA} & 4.3 & $0.79\pm 0.02$ & $0.91\pm 0.05$ & $0.84\pm 0.04$ & $0.71\pm 0.04$ & $0.82\pm 0.06$ \\
H I 1216 {\AA} (Ly$\alpha$) & 4.3 & $0.89\pm 0.02$ & $0.75\pm 0.04$ & $0.72\pm 0.04$ & $0.79\pm 0.04$ & $0.87\pm 0.06$ \\
Mg II k 2796 {\AA} & (3.9) & $0.95\pm 0.02$ & $0.91\pm 0.05$ & $0.84\pm 0.04$ & $0.86\pm 0.04$ & $0.93\pm 0.06$ \\
Mg II h 2803 {\AA} & (3.9) & $0.97\pm 0.03$ & $0.88\pm 0.05$ & $0.85\pm 0.04$ & $0.88\pm 0.04$ & $0.98\pm 0.07$ \\
Mg II k$+$h & (3.9) & $0.96\pm 0.02$ & $0.89\pm 0.05$ & $0.84\pm 0.04$ & $0.87\pm 0.04$ & $0.94\pm 0.06$ \\
Ca II K 3934 {\AA} & (3.8) & $0.87\pm 0.03$ & $1.01\pm 0.07$ & $0.90\pm 0.06$ & $0.98\pm 0.07$ & \nodata\\
Ca II H 3968 {\AA} & (3.8) & $0.86\pm 0.04$ & $1.20\pm 0.08$ & $0.98\pm 0.08$ & $0.69\pm 0.05$ & \nodata\\
H I 6563 {\AA} (H$\alpha$) & (3.8) & $-1.46\pm 0.14$ & $1.42\pm 0.17$ & $3.06\pm 1.66$ & $0.87\pm 0.19$ & \nodata\\
\enddata
\tablecomments{Column 3 shows the power-law index $\alpha$, taken from Table \ref{tab:powerlaw}. Columns 4, 5, 6, and 7 are the power-law indices that are measured for the data points in different subsets of the solar activity cycle, which are (I) the rising phase from May 2010 to August 2012; (I\hspace{-.1em}I) the solar maximum from September 2012 to July 2015; (I\hspace{-.1em}I\hspace{-.1em}I) the declining phase from August 2015 to November 2017; and (I\hspace{-.1em}V) the solar minimum from December 2017 to February 2020. There are no $\alpha_{\rm I\hspace{-.1em}V}$ for \ion{Ca}{2} K, \ion{Ca}{2} H, and H$\alpha$ because the ISS observation terminated in October 2017.}
\end{deluxetable*}


\bibliography{toriumi2021}{}
\bibliographystyle{aasjournal}



\end{document}